\documentstyle[12pt,epsfig,color]{article}
\begin{document}

\begin{center}
{\Large\bf Clustering of dark matter in interacting tachyon dark energy with $\Lambda$CDM background}
\\[15mm]
Ankan Mukherjee\footnote{Email: ankan.ju@gmail.com}
\vskip 0.5 cm
{\em $^{1}$Centre for Theoretical Physics,\\ Jamia Millia Islamia, Jamia Nagar, New Delhi-110025, India.}\\[15mm]
\end{center}

\vspace{0.5cm}
\vspace{0.5cm}
\pagestyle{myheadings}
\newcommand{\be}{\begin{equation}}
\newcommand{\ee}{\end{equation}}
\newcommand{\bea}{\begin{eqnarray}}
\newcommand{\eea}{\end{eqnarray}}

\begin{abstract}
One of the non-canonical descriptions of scalar field dark energy is the tachyon. The present work is devoted to study the dynamics of dark matter overdensity in a conformally  coupled tachyon field dark energy model. The model is tuned to mimic the $\Lambda$CDM cosmology at background level. The semi-analytic spherical collapse model of dark matter overdensity is adopted to study the  nonlinear evolution. The effects of non-minimal coupling in the energy budget on the clustering of dark matter is investigated. It is observed that the growth rate of matter overdensity is higher in presence of the non-minimal coupling. The critical density at collapse is suppressed in case of interaction. Further the number counts of dark matter halos or galaxy clusters along redshift are studied using the Press-Schechter and Sheth-Tormen halo mass functions. Suppression in the number of dark matter halos is observed when the interaction is allowed. A comparison of cluster number count in the present model and $\Lambda$CDM is carried out. The present model allowing the interaction produces much lower number of galaxy clusters compared to $\Lambda$CDM, but without interaction the cluster number count  is slightly higher than $\Lambda$CDM cluster count.    
\end{abstract}

\vskip 1.0cm

\section{Introduction}

Cosmological observations revels that the present universe is going through a phase of accelerated expansion \cite{Riess:1998cb,Perlmutter:1998np,Schmidt:1998ys,Aghanim:2018eyx}. 
Within the regime of General Relativity, cosmic acceleration can be explained by introducing an exotic component in the energy budget of the universe. This exotic component is dubbed as {\it dark energy}. There are various theoretical explanations regarding the physical entity of dark energy, starting from the vacuum energy density \cite{Carroll:2000fy,Peebles:2002gy,Padmanabhan:2002ji} to different canonical and non-canonical scalar fields \cite{Ratra:1987rm,Carroll:1998zi,Caldwell:1997ii,Padmanabhan:2002cp,Copeland:2004hq,Copeland:2006wr} or some exotic fluid with special nature of the equation of state \cite{Kamenshchik:2001cp,Bento:2002ps}. 

Though dark energy has started dominating the dynamics of the universe in the recent past (around redshift $z<0.8$), its presence in the early universe might have its imprint on the large scale structure of the universe. Hence it is important to probe the effect of different dark energy models on the formation of cosmic large scale structures. The mechanism of structure formation can be understood by exploring the nonlinear evolution of the overdense regions in the matter field. The sophisticated technique to probe the large scale structure formation is N-body simulation \cite{Maccio:2003yk,Baldi:2010vv,Boni:2011}. A semi-analytic method of probing the structure formation is the spherical collapse model \cite{Gunn:1972sv,Liddle:1993fq,Padmanabhan:1999,Nunes:2004wn,Pace:2010} of matter overdensities. The basic idea of spherical collapse model is to assume the the overdense regions as  spherical patches which will grow and eventually collapse due to gravitational pull. The dynamics of the spherical overdense region is effected by the background cosmology. Hence in this method we can probe the effects of background cosmological model on the formation of large scale structure in the universe. The evolution dynamics of matter overdensity and the number of objects, formed by its collapse, carries the signature background cosmology. With the technical developments in observational cosmology the detection of distance galaxies and the large scale structure in the universe becomes more precise. Analytic study of dark matter clustering and formation of dark matter halos in different dark energy scenario would be imperative to confront  the models with present and upcoming galaxy survey and cosmic large scale structure observations.  

In the present work, clustering of dark matter and the number count of dark matter halos are studied for an interacting tachyon dark energy scenario with $\Lambda$CDM background evolution. The semi-analytic approach of spherical collapse of matter overdensity is utilized in the present context. Extensive study of spherical collapse and large scale structure formation in different dynamical dark energy scenarios are already there in literature \cite{Pace:2010,Pace:2013pea,Delliou:2012ik,Nazari-Pooya:2016bra,Mota:2008ne,Devi:2010qp,Pace:2017qxv,Rajvanshi:2018xhf,Rajvanshi:2020das,Sapa:2018jja,Pace:2019vrs}. Mota and Bruck \cite{Mota:2004pa} have exhaustively studied the spherical collapse scenario with different quintessence potential. The imprints of coupling in the dark sector on the dynamics of matter overdense region are studied there in \cite{He:2010ta}. A comparative study of deviation of spherical collapse for different dark energy models are discussed by Nunes, Silva and Aghamin \cite{Nunes:2005fn}. Structure formation in interacting dark energy dark matter scenario has been explored through numerical simulation in \cite{Baldi:2010vv}. The semi-analytic approach, namely the spherical collapse model, for interacting dark energy dark matter scenario has been recently explored by Barros, Barreiro and Nunes \cite{Barros:2019hsk} where a time varying coupling between quintessence scalar field and the dark matter is considered. Aspects of spherical collapse in coupled dark energy cosmology are discussed by Wintergerst and Pettorino \cite{Wint:2010}. Besides canonical scalar field dark energy, the non-canonical scalar field dark energy cases are also considered in the context of the evolution of matter density contrast, spherical collapse and structure formation \cite{Setare:2017,Singh:2019bfd}. As already mentioned, in the present work, a non-canonical scalar field dark energy, namely the tachyon, interacting with dark matter has been invoked in the context of spherical collapse. The dynamics of spherical overdensity for interacting tachyon with $\Lambda$CDM background is investigated in the present analysis. As the $\Lambda$CDM cosmology is consistent with almost all the cosmological observations at background level, any viable cosmological model should produce $\Lambda$CDM like background evolution. At perturbative level, there is a small tension in the estimated values of r.m.s. fluctuation of matter power spectrum ($\sigma_8$) from Planck-$\Lambda$CDM and the value of $\sigma_8$ measured from the redshift space distortion. The evolution of the perturbation and clustering of dark matter bear the signature of dark energy properties. Thus the study of dark matter clustering is useful to distinguish dark energy models which are degenerate at background level.

Tachyon is a non-canonical description of scalar field dark energy. In the present work, the background is tuned to mimic the $\Lambda$CDM by imposing the equality of  Hubble parameter and its first derivative for the present model and the $\Lambda$CDM. Further the tachyon filed is allowed to interact with the matter field through a non-minimal coupling. Fluctuation of matter density in a coupled quintessence with $\Lambda$CDM background is explored by Barros {\it et al} \cite{Barros:2018efl}. The spherical collapse of matter overdensity in coupled quintessence with $\Lambda$CDM background is studied by Barros, Barreiro and Nunes \cite{Barros:2019hsk}. In the present analysis, the interaction between the tachyon dark energy and the dark matter is constructed from phenomenological assumption about the interaction function. The utility of fixing the background as the $\Lambda$CDM is that the scalar filed dark energy density and the equation of state parameter can be expressed independent of the scalar filed potential. Thus the analysis is independent of any specific choice of the scalar filed potential. The effect of interaction on the evolution of matter overdensity and its collapse is investigated. The number of collapsed objects or the dark matter halos along the redshift is also emphasized. The halo mass function formula is important to study the number count of dark matter halos. Two different formalism of halo mass function, namely the Press-Schechter mass function and Sheth-Tormen mass function are utilized in the present context. Both the mass functions formalisms are based on the assumption of Gaussian matter density filed. The difference between the number counts of dark matter halos in these two formalism is emphasized. Further the results are compared with the corresponding $\Lambda$CDM scenario.

The manuscript is organized as the following. In section \ref{itde}, the theoretical formulation of interacting tachyon dark energy in $\Lambda$CDM background is discussed. In section \ref{sphericalcollapse}, evolution of matter overdensity and the spherical collapse are discussed in the context of the present model. In section \ref{clustercount}, the number count of dark matter halos or galaxy clusters are studied. Finally the results are summarized in the discussion section (section \ref{discuss}).

\section{Interacting tachyon dark energy with $\Lambda$CDM background}
\label{itde}

The homogeneous and isotropic universe is described by Friedmann-Lemaitre-Robertson-Walker (FLRW) metric. In a spatially flat geometry, FLRW metric is written as,
\be
ds^2=-dt^2+a^2(t)\delta_{ij}dx^idx^j,
\ee
where $a(t)$ is the scale factor. The Hubble expansion rate or the Hubble parameter is defined as $H=\dot{a}/a$, where the overhead dot denotes the differentiation with respect to time $t$. The background evolution is governed by Friedmann equations. In terms of Hubble parameter and and its time derivative, the Friedmann equations are written as,
\be
3H^2(z)=8\pi G\left(\rho_m+\rho_{r}+\rho_{DE}\right),
\label{friedmann1}
\ee
\be
2\dot{H}+3H^2=-8\pi G(p_r+p_{de}),
\label{friedmann2}
\ee
where $\rho_m$, $\rho_r$, $\rho_{de}$ are respectively the matter, radiation and dark energy densities, and $p_r$, $p_{de}$ are pressure contributions of radiation and dark energy. In the present context, the non-canonical scalar field, namely the tachyon, is considered as the candidate of dark energy. The dark energy density $\rho_{de}$ is further denoted as the scalar field energy density $\rho_{\phi}$. Tachyon field in the context of dark energy was invoked by Padmanabhan \cite{Padmanabhan:2002cp}. Many more discussions on tachyon dark energy are there in literature \cite{Copeland:2004hq,Bagla:2002yn,Abramo:2003cp,Aguirregabiria:2004xd,Guo:2004dta,Martins:2016lgr}. Aspects of spherical collapse in tachyon dark energy have been discussed by  Rajvanshi and Bagla \cite{Rajvanshi:2020das} and by Setare, Felegary and Darabi \cite{Setare:2017}. Effect of inhomogeneous tachyon dark energy on dark matter clustering is studied by Singh, Jassal and Sharma \cite{Singh:2019bfd}. A tachyon field is described by the Lagrangian,
\be
L=-V(\phi)\sqrt{1-\partial^{\mu}\phi\partial_{\mu}\phi},
\ee 
where $V(\phi)$ is scalar field potential. In case of dark energy, the scalar field is assumed to be homogeneous and evolving with time, i.e. $\phi=\phi(t)$. The energy density ($\rho_{\phi}$) and pressure ($p_{\phi}$) of the tachyon field is given as,
\be
\rho_{\phi}=\frac{V(\phi)}{\sqrt{1-\dot{\phi}^2}}, ~~~~~  p_{\phi}=-V(\phi)\sqrt{1-\dot{\phi}^2}.
\label{rhophi_pphi}
\ee 
Consequently the equation of state parameter of the scalar field is given as, $w_{\phi}=\dot{\phi}^2-1$. The equation of state of tachyon field does not dependent on the choice of the scalar field potential. As already mentioned, in the present context a $\Lambda$CDM like background evolution is tuned. In $\Lambda$CDM cosmology, the dominant contributions to the energy density come from the vacuum energy density $\rho_{\Lambda}$ and the cold dark matter $\rho_{cdm}$. The vacuum energy density $\rho_{\Lambda}$ remains constant throughout the evolution and $\rho_{cdm}\propto a^{-3}$.  To ensure a $\Lambda$CDM like background in the present model, the Hubble parameter and its first order time derivative for the present model are taken to be equal to that of $\Lambda$CDM. The equalities relate the energy densities as,
\be
\rho_{cdm}+\rho_{\Lambda}=\rho_{m}+\rho_{\phi},
\label{rho_relations1}
\ee
and
\be
\rho_{cdm}=\rho_m+\rho_{\phi}+p_{\phi}.
\label{rho_relations2}
\ee
Utilizing equation (\ref{rhophi_pphi}), (\ref{rho_relations1}) and (\ref{rho_relations2}), the scalar field potential is expressed as 
\be
V(\phi)=\frac{\rho_{\Lambda}}{\sqrt{1-\dot{\phi}^2}}.
\label{Vphi}
\ee
Similarly the matter density is expressed as,
\be
\rho_m=\rho_{cdm}-\frac{\rho_{\Lambda}\dot{\phi}^2}{(1-\dot{\phi}^2)}.
\label{rhom}
\ee
Thus physical quantities of a tachyon dark energy model are expressed in terms of $\phi$, $\dot{\phi}$ and $\rho_{\Lambda}$ and $\rho_{cdm}$ in the present context. Further, interaction between the scalar field and the dark matter is introduced. The total conservation of the energy momentum tensor ($T^{\mu\nu}$) is obtained from the contracted Bianchi identity $T^{\mu\nu}_{;\mu}=0$. For homogeneous distribution of different components in FLRW geometry, the conservation equation of individual components are written as,
\be
\dot{\rho}_m+3H\rho_m=-Q,
\label{mat_con}
\ee 
\be
\dot{\rho}_{de}+3H(1+w)\rho_{de}=Q,
\label{de_con}
\ee
\be
\dot{\rho}_r+4H\rho_r=0.
\label{r_con}
\ee
The $Q$ is the interaction function that determines characteristics of energy transfer between the dark matter and dark energy component. Independent conservation of radiation energy density is assumed in the present analysis (equation \ref{r_con}). The conservation of $\rho_{\phi}$ yields the evolution equation of the scalar field as,
\be
\ddot{\phi}+(1-\dot{\phi}^2)\left[3H\dot{\phi}+\frac{1}{V}\frac{dV}{d\phi}\right]=\frac{(1-\dot{\phi}^2)^{3/2}}{V\dot{\phi}}Q.
\label{phi_eq}
\ee
From equation (\ref{Vphi}) the differentiation of the scalar field potential $V(\phi)$ with respect to $\phi$ is expressed as,
\be
\frac{1}{V}\frac{dV}{d\phi}=\frac{\ddot{\phi}}{(1-\dot{\phi}^2)}.
\ee
Finally equation (\ref{phi_eq}) can be written in the form which is independent of the scalar filed potential,
\be
2\ddot{\phi}+3H(1-\dot{\phi}^2)\dot{\phi}=\frac{(1-\dot{\phi}^2)^{2}}{\rho_{\Lambda}\dot{\phi}}Q.
\label{phi_eq2}
\ee

%%%%%%%%%%%%%%%%%%%%%
\begin{figure}[tb]
\begin{center}
\includegraphics[angle=0, width=0.38\textwidth]{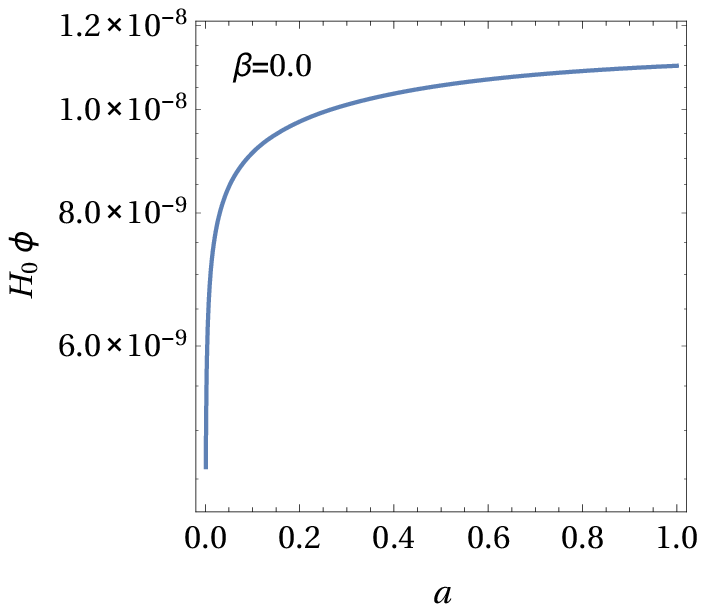}
\includegraphics[angle=0, width=0.34\textwidth]{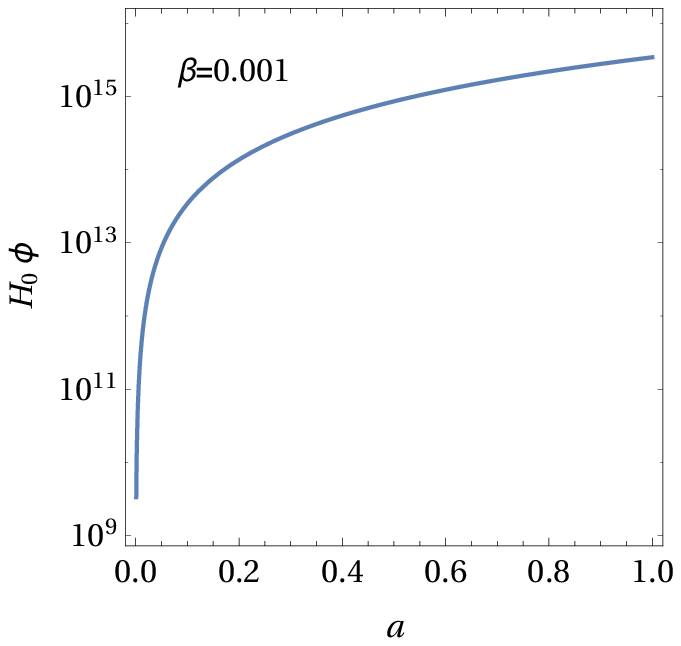}
\end{center}
\caption{{\small Plots of the redefined scalar field $\tilde{\phi}=H_0\phi$  as a function of the scale factor $a$ obtained from the numerical solution of equation (\ref{phi_a3})  with initial conditions, fixed at $a_i=10^{-5}$, as $\tilde{\phi}_i=0.0$, $\tilde{\phi}'_i=0.0001$. The left panel shows the curve for $\beta=0.0$ and the right panel shows the curve  for $\beta=0.001$.}}
\label{Phia}
\end{figure}
%%%%%%%%%%%%%%%%%%%%%%%%%%%%%%%%%%%%%%%%

The interaction function effects the evolution of the scalar field. There is no theoretical compulsion about the choice of the interaction function. Thermodynamical requirement demands that the energy transfer in the interaction  should be from  dark matter to dark energy \cite{Pavon:2007gt}. To satisfy the condition of energy flow, the interaction function $Q$ should be positive according to the choice of signature in equations (\ref{mat_con}) and (\ref{de_con}).    The interaction dynamics of dark energy and dark matter is  usually studied using phenomenological assumption regarding the interaction function. In the present analysis,  the form of the interaction functions is assumed as,
\be
Q=\beta H\rho_{m}.
\label{Q}
\ee
The interaction function is proportional to the dark matter density $\rho_m$ and interaction rate is linear to the Hubble expansion rate.
The dimensionless parameter $\beta$ determines the strength of coupling between the scalar field and the matter field.
Changing the argument of differentiation to the scale factor $a$, the scalar field equation (\ref{phi_eq2}) is obtained as, 
\be
\phi''+\left(\frac{1}{a}+\frac{H'}{H}\right)\phi'+\frac{3\phi'}{2a}(1-a^2H^2\phi'^2)=\frac{(1-a^2H^2\phi'^2)^{2}}{2\rho_{\Lambda}a^3H^3\phi'}Q.
\label{phi_a}
\ee
For the interaction function, given is equation (\ref{Q}), equation (\ref{phi_a}) yields as,
\be
\phi''=-\left(\frac{1}{a}+\frac{H'}{H}\right)\phi'-\frac{3\phi'}{2a}(1-a^2H^2\phi'^2)+\frac{\beta(1-a^2H^2\phi'^2)^{2}}{2a^3H^2\phi'}  \left(\frac{\rho_m}{\rho_{\Lambda}}\right).
\label{phi_a2}
\ee
Finally using the expression of $\rho_m$ from equation (\ref{rhom}), the equation is written as,
\be
\phi''=-\left(\frac{1}{a}+\frac{H'}{H}\right)\phi'-\frac{3\phi'}{2a}(1-a^2H^2\phi'^2)+\frac{\beta(1-a^2H^2\phi'^2)^{2}}{2a^3H^2\phi'}  \left(\frac{\rho_{cdm0}}{\rho_{\Lambda}}a^{-3}-\frac{a^2H^2\phi'^2}{1-a^2H^2\phi'^2}\right),
\label{phi_a2}
\ee
where $\rho_{cdm0}$ is the present density of cold dark matter in $\Lambda$CDM cosmology. For convenience, the scalar field is redefined in a dimensionless way as $\tilde{\phi}=\phi H_0$, where $H_0$ is the present Hubble parameter or the Hubble constant. The Hubble parameter $H(a)$ is rescaled by $H_0$ as $h(a)=H(a)/H_0$. In terms of the redefined scalar field, equation (\ref{phi_a2}) can be written as,
\be
\tilde{\phi}''=-\left(\frac{1}{a}+\frac{h'}{h}\right)\tilde{\phi}'-\frac{3\tilde{\phi}'}{2a}(1-a^2h^2\tilde{\phi}'^2)+\frac{\beta(1-a^2h^2\tilde{\phi}'^2)^{2}}{2a^3h^2\tilde{\phi}'}  \left(\frac{\rho_{cdm0}}{\rho_{\Lambda}a^3}-\frac{a^2h^2\tilde{\phi}'^2}{1-a^2h^2\tilde{\phi}'^2}\right).
\label{phi_a3}
\ee
Equation (\ref{phi_a3}) is studied numerically to obtain the evolution of the scalar field. The initial conditions are fixed at the scale factor $a=10^{-5}$ and the initial conditions are $\tilde{\phi}_i=0$ and $\tilde{\phi}'_i=0.0001$. Figure \ref{Phia} shows the curves of the scalar field $\tilde{\phi}(a)$ for different values of the coupling parameter $\beta$. When the interaction coupling parameter is fixed to $\beta=0$, that means no interaction is allowed, the scalar field shows a slowly varying nature. On the other hand, non zero coupling parameter allows a rapid evolution of the scalar field. The value of the ratio $(\rho_{cdm0}/\rho_{\Lambda})$ is fixed from the Planck-$\Lambda$CDM estimation by Planck 2018 \cite{Aghanim:2018eyx}.

%%%%%%%%%%%%%%%%%%%%%
\begin{figure}[tb]
\begin{center}
\includegraphics[angle=0, width=0.34\textwidth]{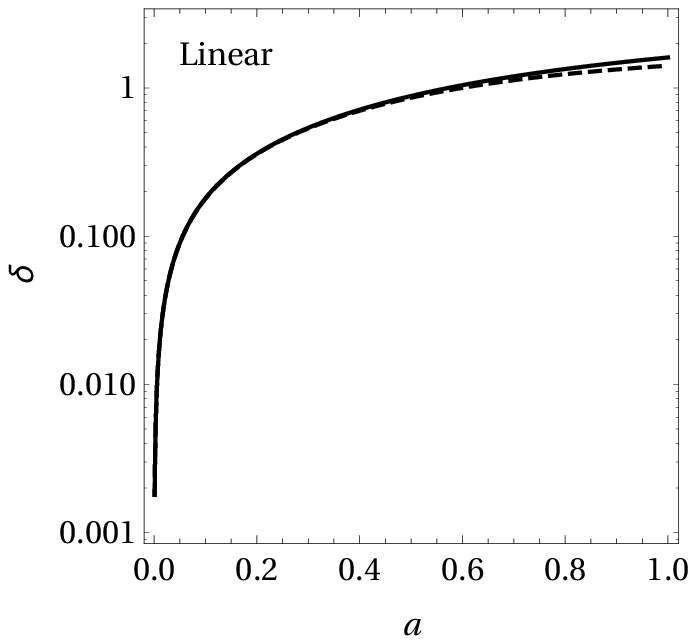}
\includegraphics[angle=0, width=0.34\textwidth]{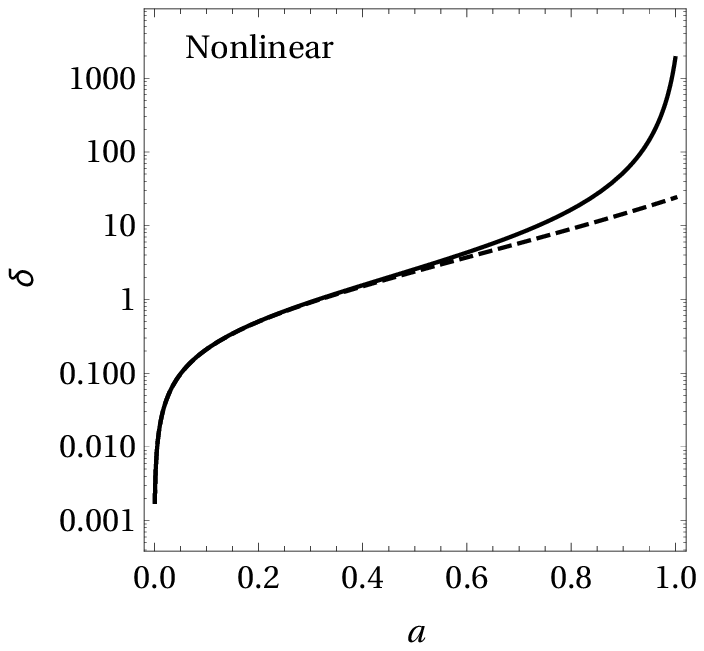}
\end{center}
\caption{{\small Linear and nonlinear evolution of $\delta(a)$ obtained from the numerical solutions of equation (\ref{dela_nonlin}) and (\ref{dela_lin}). The boundary conditions are fixed at $a=10^{-5}$ as $\delta_i=3.0\times 10^{-5}$ and $\delta'_i=0.0$. The solid curves are obtained for $\beta=0.001$ and the dashed curves are obtained for $\beta=0.0$.}}
\label{delMplot}
\end{figure}
%%%%%%%%%%%%%%%%%%%%%%%%%%%%%%%%%%%%%%%%

%%%%%%%%%%%%%%%%%%%%%
\begin{figure}[tb]
\begin{center}
\includegraphics[angle=0, width=0.35\textwidth]{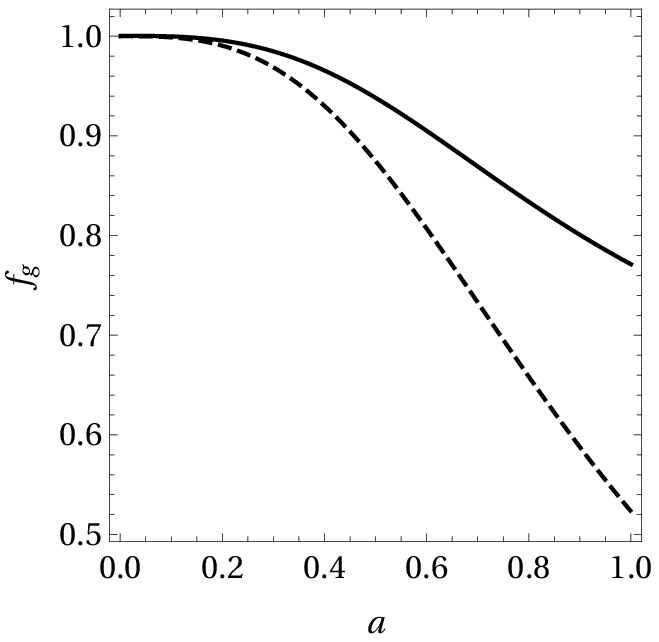}
\includegraphics[angle=0, width=0.35\textwidth]{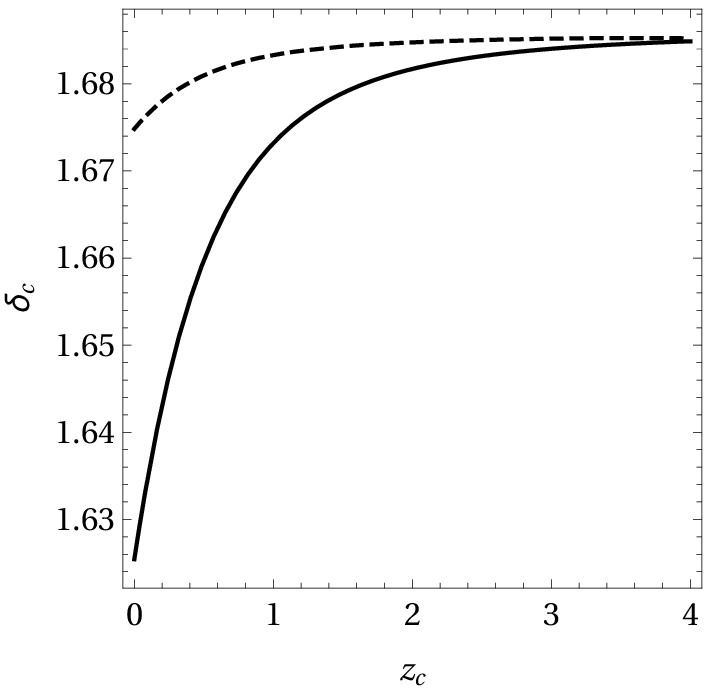}
\end{center}
\caption{{\small Left panel shows the linear growth rate $f_g$ as a function of the scale factor. Right panel shows the critical density at collapse $\delta_c$ as a function of the redshift at collapse $z_c$. The solid curves are obtained for $\beta=0.001$ and the dashed curves are obtained for $\beta=0.0$. }}
\label{fig:delC_beta}
\end{figure}
%%%%%%%%%%%%%%%%%%%%%%%%%%%%%%%%%%%%%%%%

\section{Evolution of matter overdensity and spherical collapse}
\label{sphericalcollapse}

In this section, the evolution of matter overdensity in the present dark energy model is discussed. The evolution equation of matter overdensity is conveniently written in terms of the matter density contrast, defined as $\delta=\frac{\Delta\rho_m}{\rho_m}$ where $\Delta\rho_m$ is the deviation from homogeneous matter density $\rho_m$. The overdense region initially expands with Hubble expansion. At the same time it gathers mass due to gravitational attraction. After certain amount of mass accumulation, the overdense region stops the expansion and starts to collapse. The collapse of the overdense region is the fundamental process of large scale structure formation in the universe. It is essential to study the nonlinear evolution of the matter overdensities to understand the dynamics of the structure formation.  Spherical collapse model \cite{Gunn:1972sv,Liddle:1993fq,Padmanabhan:1999,Nunes:2004wn} is the simplest approach to probe the evolution of the matter density contrast at the nonlinear regime. It is a semi-analytic approach that assumes the overdense  regions are spherically symmetric and the density inside the sphere is homogeneous.  The nonlinear differential equation of matter density contrast in case of interacting dark energy is discussed in \cite{Pace:2013pea,Wint:2010,Savastano:2019zpr}.  For the present interaction function the equation is written as,  
\be
\ddot{\delta}+\dot{\delta}(2-\beta)H-4\pi G\rho_m\delta(1+\delta)(1+2\beta^2)-\frac{4}{3}\frac{\dot{\delta}^2}{1+\delta}=0,
\label{del_nonlin}
\ee
and the linear version of equation (\ref{del_nonlin}) is given as,
\be
\ddot{\delta}+\dot{\delta}(2-\beta)H-4\pi G\rho_m\delta(1+2\beta^2)=0.
\label{del_lin}
\ee
As the background is fixed to evolve like $\Lambda$CDM, the matter density $\rho_m$ in equation (\ref{del_nonlin}) and (\ref{del_lin}) is expressed by the relation given in equation (\ref{rhom}). Thus the scalar field evolution effects the dynamics of $\delta$ in this context. Taking the scale factor $a$ as the argument of differentiation, equation (\ref{del_nonlin}) is obtained as,
\be
\delta''+\left(\frac{3-\beta}{a}+\frac{h'}{h}\right)\delta'-\frac{3}{2}\left(\frac{\Omega_{cdm0}}{a^5h^2}-\frac{(1-\Omega_{cdm0})\tilde{\phi}'^2}{(1-a^2h^2\tilde{\phi}'^2)}\right)\delta(1+\delta)(1+2\beta^2)-\frac{4}{3}\frac{\delta'^2}{1+\delta}=0,
\label{dela_nonlin}
\ee
and similarly the linear equation (eq. \ref{del_lin}) is obtained as,
\be
\delta''+\left(\frac{3-\beta}{a}+\frac{h'}{h}\right)\delta'-\frac{3}{2}\left(\frac{\Omega_{cdm0}}{a^5h^2}-\frac{(1-\Omega_{cdm0})\tilde{\phi}'^2}{(1-a^2h^2\tilde{\phi}'^2)}\right)\delta(1+2\beta^2)=0.
\label{dela_lin}
\ee
Equation (\ref{dela_nonlin}) and (\ref{dela_lin}) are studied numerically. The linear and nonlinear evolution of $\delta$ are shown in figure \ref{delMplot} for different values of $\beta$. In figure \ref{delMplot} the initial conditions are fixed at $a=10^{-5}$ as $\delta_i=0.00003$ and $\delta'_{i}=0.0$. The linear and  nonlinear evolutions are similar at early time. But at later stage, the nonlinear growth rate is higher than the linear growth rate and eventually the nonlinear evolution leads to the singularity of $\delta$. The singularity of $\delta$ indicates the collapse of the matter overdensity. The linear evolution is less effected by the change of coupling parameter $\beta$. But the effect of interaction is more prominent in case of nonlinear evolution. With the same boundary conditions, the nonliear $\delta$ grows at faster rate in case of nonzero $\beta$. Linear growth function is defined as $g(a)=\delta(a)/\delta_0$, where $\delta_0$ is the present value of linear density contrast. The linear growth rate is defined as $f_g=\frac{d\ln{\delta}}{d\ln{a}}$.  The linear growth function is obtained from the solution of equation (\ref{dela_lin}). The right panel of figure \ref{fig:delC_beta} shows the $f_g(a)$ curves for $\beta=0.0$ and $\beta=0.001$. The linear growth rate is higher in case of nonzero $\beta$ at latetime. 
 
Another important quantity in the study of dark matter clustering in a spherically collapsing scenario is the critical density contrast at collapse ($\delta_c$). It is defined as the value of the linear density contrast at the redshift where the nonlinear density contrast diverges. Changing the initial value of $\delta$ in the differential equation (equation \ref{dela_nonlin}), the redsift of nonlinear collapse is changed. The initial value of $\delta$ required to have a collapse at $z=0$ is $\delta_i=3.0412\times 10^{-5}$ for $\beta=0.001$, and $\delta_i=3.5593\times 10^{-5}$ for $\beta=0.0$. The $\delta_c$ is determined from the linear equation of $\delta$ (equation (\ref{dela_lin})) with the same initial condition. The curves of $\delta_c(z)$ for the present model are shown in the left panel of figure \ref{fig:delC_beta}. The critical density at collapse $\delta_c(z)$ is essential to study the number of collapsed objects or dark matter halos along the redshift.

\section{Halo mass function and cluster number count}
\label{clustercount}

In this section, the number count of collapsed object along the redshift is studied for the present model. The collapsed objects are called the dark matter halos. The distribution of ordinary baryonic  matter follows the distribution of dark matter due to gravitational attraction. The clusters of galaxies are actually embedded in dark matter halos. Thus the distribution of dark matter halos can be tracked by observing the distribution of galaxy clusters. There are two different mathematical formulations of halo mass to evaluate the number count of collapsed objects or halos along the redshift. The first one is the Press-Schechter mass function formalism \cite{Press:1973iz}. Later a generalization of Press-Shechter mass function was proposed by Sheth and Tormen \cite{Sheth:1999mn}. Both these formalism stand up with the assumption of a Gaussian distribution of the matter density field. In the present analysis, both of these mass functions are utilized. The comving number density of collapsed objects or dark matter halos at a certain redshift $z$ having mass range $M$ to $M+dM$ is expressed as,
\be
\frac{dn(M,z)}{dM}=-\frac{\rho_{m0}}{M}\left(\frac{d}{dM}\ln{\sigma(M,z)}\right)f(\sigma(M,z)),
\ee
where $f(\sigma)$ is the {\it mass function}. The mathematical formulation of the mass function was first proposed by Press and Schechter \cite{Press:1973iz}, which is given as,
\be
f_{PS}(\sigma)=\sqrt{\frac{2}{\pi}}\frac{\delta_c(z)}{\sigma(M,z)}\exp{\left[-\frac{\delta^2_c(z)}{2\sigma^2(M,z)} \right]}.
\label{fPS}
\ee
The $\sigma(M,z)$ is the corresponding rms density fluctuation in a sphere of radius $r$ enclosing a mass M. This can be expressed in terms of the linearized growth factor $g(z)=\delta{z}/\delta(0)$, and the rms of density fluctuation at a fixed length $r_8=8h_0^{-1}$Mpc as,

\be
\sigma(z,M)=\sigma(0,M_8)\left(\frac{M}{M_8}\right)^{-\gamma/3}g(z),
\ee
where $M_8=6\times10^{14}\Omega_{m0}h_0^{-1}M_{\odot}$, the mass within a sphere of radius $r_8$ and the $M_{\odot}$ is  the solar mass and $h_0$ is the Hubble constant $H_0$ scaled by $100km.s^{-1}Mpc$. The $\gamma$ is defined as
\be
\gamma=(0.3\Omega_{m0}h_0+0.2)\left[2.92+\frac{1}{3}\log{\left(\frac{M}{M_8}\right)}\right].
\ee
Finally the number of collapsed objects within a mass range $M_i<M<M_s$ per redshift per square degree yield as,
\be
{\mathcal N}(z)=\int_{1deg^2}d\Omega\left( \frac{c}{H(z)}\left[\int_0^z\frac{c}{H(x)}dx\right]^2\right)\int_{M_i}^{M_s}\frac{dn}{dM}dM.
\label{numbercount_eq}
\ee
This ${\mathcal N}(z)$ is called the number count of dark matter halos or cluster number count. Press-Schechter mass function formula is successful to depict a general nature of cluster number count. But it predicts higher abundance of galaxy cluster at low redshift and lower abundance  at high redshift compared to the result obtained in simulation of dark matter halo formation \cite{Reed:2006rw}. To alleviate this issue, a modification is proposed by Sheth and Tormen \cite{Sheth:1999mn}, which is given as,
\be
f_{ST}(\sigma)=A\sqrt{\frac{2}{\pi}}\left[1+\left( \frac{\sigma^2(M,z)}{\delta^2_c(z)}\right)^p\right]\frac{\delta_c(z)}{\sigma(M,z)}\exp{\left[-\frac{a\delta_c^2(z)}{2\sigma^2(M,z)}\right]}.
\label{fST}
\ee
The Sheth-Tormen mass function formula, given in equation (\ref{fST}), introduces three new parameters $(a,p,A)$.  For the values $(1,0,\frac{1}{2})$ of the set of parameters $(a,p,A)$ the Sheth-Torman mass funtion actually becomes the Press-Schechter mass function. In the present work, while studying the cluster number count  using Sheth-Tormen mass function formula, the values of the parameters $(a,p,A)$ are fixed at $(0.707,0.3,0.322)$ as suggested by the simulation  of dark matter clustering \cite{Reed:2006rw}.

%%%%%%%%%%%%%%%%%%%%%
\begin{figure}[tb]
\begin{center}
\includegraphics[angle=0, width=0.4\textwidth]{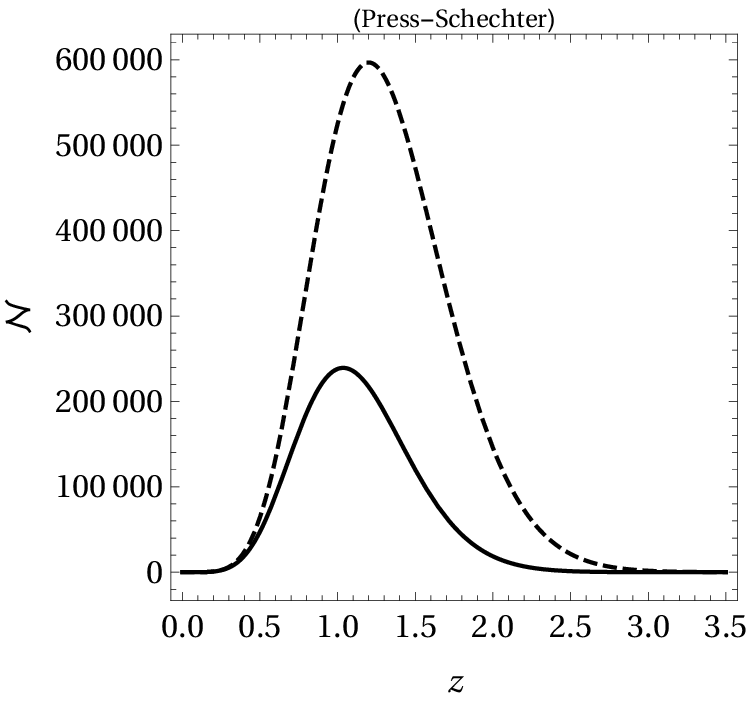}
\includegraphics[angle=0, width=0.4\textwidth]{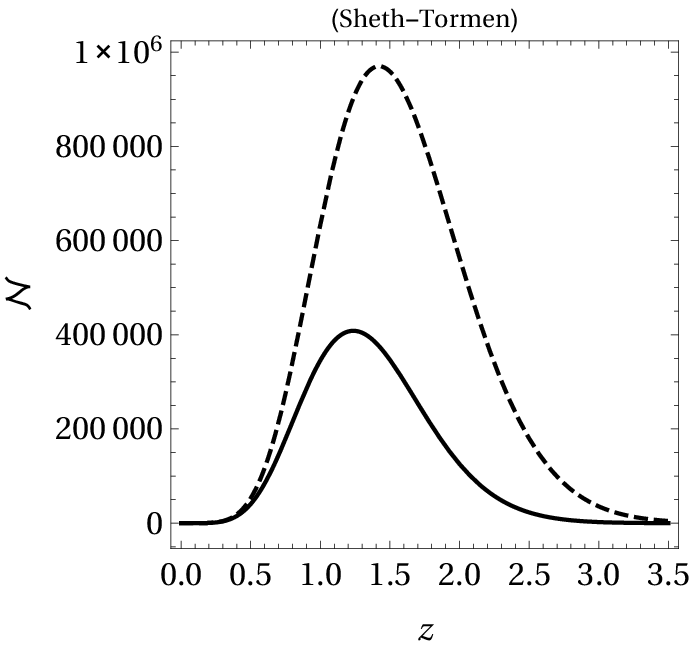}
\end{center}
\caption{{\small Plots show the cluster number count along redshift for the present dark energy model. The left panel is obtained for the Press-Schechter mass function and the right panel is obtained for Sheth-Tormen mass function. The solid curves are for $\beta=0.001$ and the dashed curves are for $\beta=0$.}}
\label{Nplot_PSST}
\end{figure}
%%%%%%%%%%%%%%%%%%%%%%%%%%%%%%%%%%%%%%%%

In figure \ref{Nplot_PSST} the plots show the cluster number count ${\mathcal N}(z)$ using Press-Schechter and Sheth-Tormen mass function for the present dark energy scenario. Values of cosmological parameters $\Omega_{m0}$, $H_0$ and $\sigma_8$ are fixed at the best fit of latest measurements from Planck-$\Lambda$CDM along with CMB lensing and BAO data \cite{Aghanim:2018eyx}. The values are $\Omega_{m0}=0.3111$, $H_0=67.66~km.s^{-1}Mpc^{-1}$ and $\sigma_8=0.8102$. The mass range is fixed as $10^{14}h^{-1}_0M_{\odot}<M<10^{16}h^{-1}_0M_{\odot}$. In figure \ref{Nplot_PSST}, plots are obtained for two different values of the coupling parameter $\beta=0.001$ and $\beta=0$. The cluster number count is highly suppressed when the interaction is allowed. 

\begin{figure}[tb]
\begin{center}
\includegraphics[angle=0, width=0.4\textwidth]{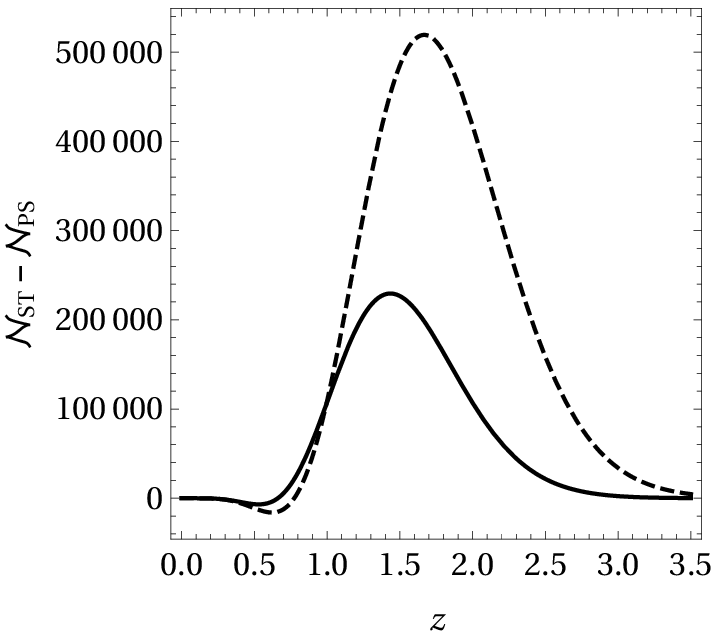}
\includegraphics[angle=0, width=0.4\textwidth]{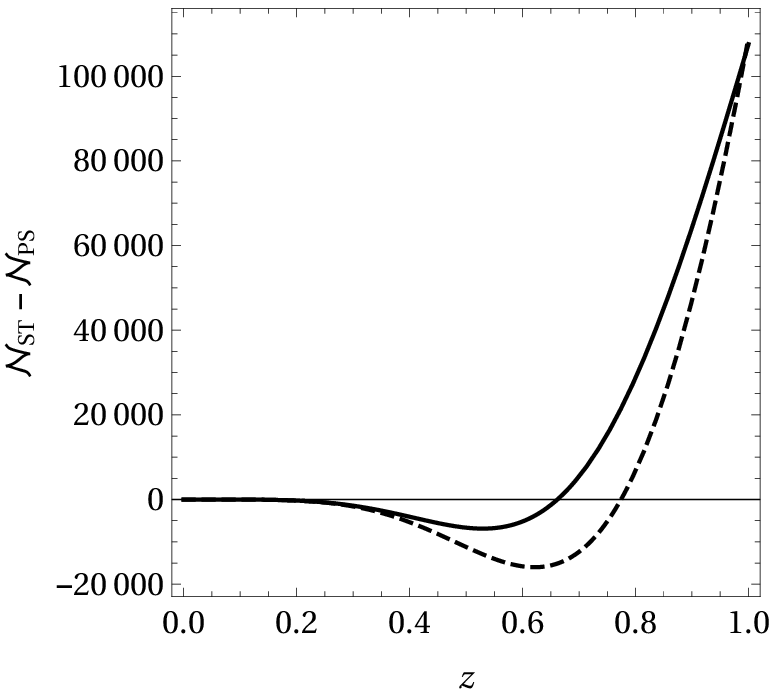}
\end{center}
\caption{{\small Plots show the difference between the cluster count obtained for Press-Schechter mass function and Sheth-Tormen mass function. The solid curves are for $\beta=0.001$ and the dashed curves are for $\beta=0$. The right panel shows the same in the redshift range $0\leq z\leq 1$.}}
\label{fig:delN_PSST}
\end{figure}

The difference between Press-Schechter and Sheth-Tormen cluster number count is shown in figure \ref{fig:delN_PSST}. The Sheth-Tormen mass function produces much higher number of dark matter halos at high redshift compared to the number of halos produced in case of Press-Schechter. But at low redshift ($z<0.8$), the Press-Schechter cluster number is slightly higher than that of Sheth-Tormen.

%%%%%%%%%%%%%%%%%%%%%
%%%%%%%%%%%%%%%%%%%%%%%%%%%%%%%%%%%%%%%%

%%%%%%%%%%%%%%%%%%%%%
\begin{figure}[tb]
\begin{center}
\includegraphics[angle=0, width=0.4\textwidth]{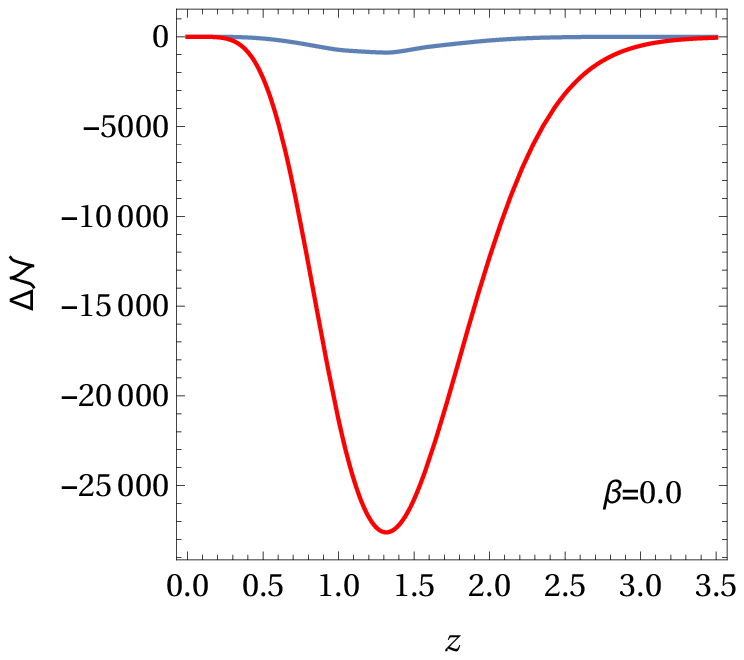}
\includegraphics[angle=0, width=0.4\textwidth]{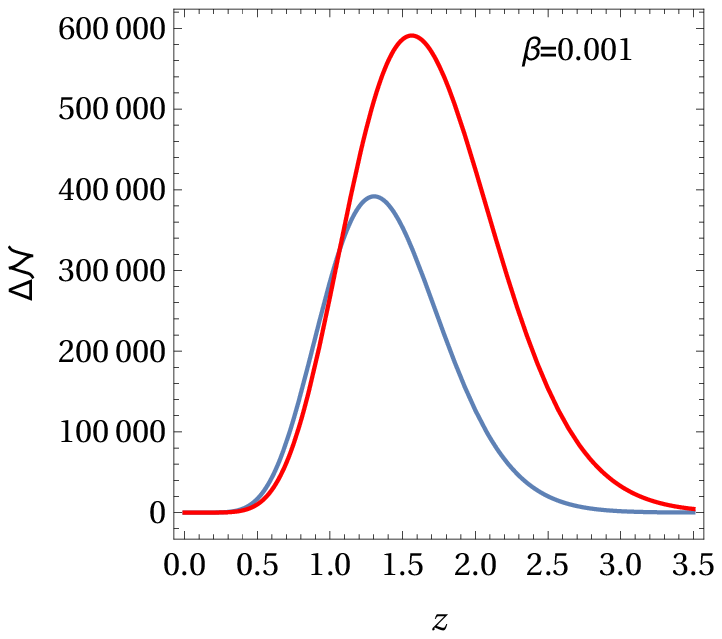}
\end{center}
\caption{{\small Plots show the difference in cluster number count for the present dark energy scenario and the $\Lambda$CDM, $\Delta{\mathcal N}=({\mathcal N}_{\Lambda CDM}-{\mathcal N}_{ITach})$. The left panel is for $\beta=0$ and the right panel is for $\beta=0.001$. The blue curves are obtaibed for Press-Schechter mass function and the red curves are obtained for Sheth-Tormen mass function.}}
\label{fig:delN_LCDM}
\end{figure}
%%%%%%%%%%%%%%%%%%%%%%%%%%%%%%%%%%%%%%%%

The cluster number count obtained in the present dark energy scenario is compared with the same for $\Lambda$CDM cosmology. The difference between the number count of dark matter halos or galaxy clusters in the present model (ITach) and $\Lambda$CDM is shown in figure \ref{fig:delN_LCDM} where the $\Delta{\mathcal N}={\mathcal N}_{\Lambda CDM}-{\mathcal N}_{ITach}$. The ${\mathcal N}_{\Lambda CDM}$ is obtained for same values of cosmological parameters $\Omega_{m0}$, $H_0$ and $\sigma_8$. In case of $\beta=0$, the number count in the present model is slightly higher than the $\Lambda$CDM. On the other hand, in case of $\beta=0.001$, the cluster number is substantially suppressed compared to the $\Lambda$CDM results. The difference of cluster number is higher in case of Sheth-Tormen mass function compared to that in Press-Schechter mass function.

\section{Discussion}
\label{discuss}
In the present work, the nonlinear evolution of matter overdensity is studied in a non-minimally coupled tachyon scalar field dark energy with $\Lambda$CDM background. The nonlinear clustering of dark matter is explored with the assumption of spherical collapse model of matter overdensity.
It is observed that the non-minimal coupling causes a faster growth of matter density contrast compared to the noninteracting scenario in the nonlinear regime (figure \ref{delMplot}). The linear growth rate is also higher at low redshift in case of interaction (left panel of figure \ref{fig:delC_beta}). In noninteracting scenario, the density contrast evolve for longer time before collapse. This causes a higher value of critical density at collapse ($\delta_c$) in case of noninteracting scenario than the interacting scenario (right panel of figure \ref{fig:delC_beta}). 

Further the number count of collapsed objects or the galaxy clusters along redshift is studied for two different mass function formalisms. It is observed that the cluster number count is substantially suppressed in case of interaction for both the mass functions (figure \ref{Nplot_PSST}). The cluster number is higher at high redshit in case of Sheth-Tormen mass function compared to the cluster count for Press-Schechter mass function. At low redshift ($z<0.8$), Press-Schechter mass function produces higher cluster number (figure \ref{fig:delN_PSST}). The difference between Press-Schechter and Sheth-Torman number count is higher in noninteracting scenario.     

Further the cluster number count in the present dark energy scenario is compared to the cluster number count in $\Lambda$CDM (figure \ref{fig:delN_LCDM}). In case of non-minimal coupling in tachyon field, the cluster number count is substantially lower that that of $\Lambda$CDM. In case of non-interacting tachyon, the cluster number count is slightly higher than that of $\Lambda$CDM, though the difference is much lower than the interacting tachyon case. The difference is higher in case of Sheth-Tormen mass function than the Press-Schechter mass function.

It worth mention at this point that the cluster number count is highly effected by the fiducial cosmological model and values of the cosmological parameters. As we assume a $\Lambda$CDM background evolution, the volume element present in the expression of the cluster number count (equation \ref{numbercount_eq}) remains invariant in the present analysis and the parameters values are fixed at the corresponding $\Lambda$CDM values. The present analysis clearly suggests that the study of nonlinear evolution and spherical collapse of matter overdensity is an useful way to distinguish various cosmological models which are degenerate at background and linear perturbation level. Exclusive study in this direction can also be implemented in purely numerical approach using N-body simulation technique. But the semi-analytic approach of spherical collapse is easier to implement in various dark energy scenario compared to the fully simulation approach.

\vskip 1.0 cm

\section*{Acknowledgment}
The author would like to acknowledge the financial support from the Science and Engineering Research Board (SERB), Department of Science and Technology, Government of India through National Post-Doctoral Fellowship (NPDF, File no. PDF/2018/001859). The author would like to thank Prof. Anjan A. Sen for useful discussions and suggestions.

\end{document}